%% file: eprint_dpf2013.tex
\documentclass[12pt]{article}
\usepackage{graphicx}
\usepackage{hyperref}
\usepackage{amsmath}
\usepackage[parfill]{parskip}
\usepackage{multirow}
\usepackage{tabularx}

\newcommand\pubnumber{DPF2013-156}
\newcommand\pubdate{\today}

\def\support{\footnote{Corresponding author}}

\def\Title#1{\begin{center} {\Large #1 } \end{center}}
\def\Author#1{\begin{center}{ \sc #1} \end{center}}

\newcommand\pubblock{\rightline{\begin{tabular}{l} \pubnumber\\
         \pubdate  \end{tabular}}}
\newenvironment{Abstract}{\begin{quotation}  }{\end{quotation}}
\newenvironment{Presented}{\begin{quotation} \begin{center} 
             PRESENTED AT\end{center}\bigskip 
      \begin{center}\begin{large}}{\end{large}\end{center} \end{quotation}}
\def\Acknowledgments{\bigskip  \bigskip \begin{center} \begin{large}
             \bf ACKNOWLEDGMENTS \end{large}\end{center}}

\input econfmacros.tex

\begin{document}
\begin{titlepage}
\pubblock

\vfill
\Title{Exotic Higgs searches in CMS}
\vfill
\Author{A.~Castaneda\support$^a$\\ 
\footnotesize $^a$Texas A\&M University at Qatar, Doha, Qatar \normalsize \\
\vspace{0.2in}
On behalf of the CMS Collaboration}
\vfill
\begin{Abstract}
We present some of the most recent results from the CMS Collaboration on searches for Higgs-like particles in models beyond the Standard Model. 
Several rare and exotic decay modes of the Higgs boson are explored. The results of the searches are relevant 
for establishing whether the 125~GeV particle observed in Higgs boson searches at the LHC has the properties 
expected for a Standard Model Higgs boson.
\end{Abstract}
\vfill
\begin{Presented}
DPF 2013\\
The Meeting of the American Physical Society\\
Division of Particles and Fields\\
Santa Cruz, California, August 13--17, 2013\\
\end{Presented}
\vfill
\end{titlepage}
\def\thefootnote{\fnsymbol{footnote}}
\setcounter{footnote}{0}

\section{Introduction}

In 2012 the ATLAS and CMS collaborations reported the observation of a new particle 
with a mass of 125~GeV~\cite{higgs}, the excess of events over the expected background corresponds 
to a local significance of more than five standard deviations, such excess is most 
significant in the two decay modes with the best mass resolution ($\gamma\gamma$ and $ZZ$). The 
observed particle and its properties are consistent with the 
long-sought Higgs boson~\cite{higgs_paper}. Following the excitement after the discovery the 
collaboration is facing a new challenge which is to reveal the true identity of this particle, 
whether it corresponds to the predicted Standard Model Higgs boson or it is part of an extended 
theory in which there could be more than one Higgs-like particle, for instance, in the context of Supersymmetry. 
 Most of these searches are undergoing and some of the results from the 2011 and 2012 CMS data taking are reviewed in this paper.

\section{The CMS experiment}

The central feature of the CMS detector is a superconducting solenoid of 6m internal diameter, 
providing a magnetic field of 3.8 T. Within the field volume, the inner tracker is formed
by a silicon pixel and strip tracker. It measures charged particles within the pseudo-rapidity
range $|\eta|<$ 2.5. The pseudo-rapidity is defined as $\eta =  \ln(\tan(\theta/2))$ and $\theta$ is the polar angle,
while $\phi$ is the azimuthal angle in radians. The tracker provides an impact parameter resolution of approximately 15~mm and 
a resolution on transverse momentum ($p_{T}$) of about 1.5\%
for 100 GeV particles. Also inside the field volume are a crystal electromagnetic calorimeter
and a brass/scintillator hadron calorimeter. Muons are measured in gas-ionization detectors
embedded in the iron flux return yoke, in the pseudo-rapidity range $|\eta|<$ 2.4, with detector
planes made using three technologies: drift tubes, cathode-strip chambers, and resistive-plate
chambers. Matching muons to tracks measured in the silicon tracker results in a transverse momentum resolution 
between 1\% and 5\%, 
for $p_{T}$ values up to 1~TeV. Extensive forward calorimetry complements the coverage provided by the barrel and endcap detectors. A more detailed
description of the CMS detector can be found in Ref.~\cite{cms}

\section{MSSM searches in CMS}

The Standard Model has provided a remarkably successful description of presently known phenomena, 
the experimental frontier has advanced into the~TeV scale with no unambiguous hints 
of additional structure so far, it seems clear that the Standard Model has to be extended 
to describe the physics at higher energies, for instance, at the reduced Planck scale ($M_{P} = (8\pi G_{Newton})^{-1/2} =2.4\times 10^{18}$~GeV) 
where the quantum gravitational effects become important. The enormous order of magnitude difference between the electroweak and Planck scales 
(commonly known as the "hierarchy problem") does not represent a problem 
itself to the Standard Model but rather a disturbing sensitivity of the Higgs potential to new physics in almost any imaginable extension 
of the Standard Model. The Standard Model demands a non-vanishing vacuum expectation value for the Higgs field, this requirement in addition to the measured properties of the weak interactions lead to a squared value of the Higgs mass ($m_{H}^{2}$) to be of the order 
of $\sim$(100~GeV)$^{2}$, the problem comes from the enormous quantum corrections to this quantity from the 
virtual effects of every particle that couples to the Higgs field.  If the scale of these corrections is of the order of $M_{p}$ the squared value of the Higgs mass is about thirty order of magnitudes larger than the required by the Standard Model.
By introducing a new symmetry relating fermions and bosons those large corrections can be effectively cancelled, 
such symmetry is commonly known as Supersymmetry (SUSY)
and it is a transformation that turns a bosonic state into a fermionic state and viceversa.
The Minimal Supersymmetric Standard Model (MSSM)~\cite{mssm} extend the number of particles in the Standard Model including the Higgs sector by adding an additional Higgs doublet giving rise to three neutral Higgs bosons ($h^{0}, H^{0},A^{0}$) and two charged ones ($H^{\pm}$).
The phenomenology of MSSM can be effectively described using only two parameters: the mass of the CP-odd Higgs boson ($m_{A}$) and the 
ratio of the vacuum expectation values of the two Higgs doublets ($\tan{\beta}$). 
\newline

\noindent
The CMS collaboration has conducted several searches in the context of the MSSM scenario, one of the peculiarities of 
this model is that for large $tan(\beta)$ the Higgs boson decays to $\tau$'s and b-quarks play a more important 
role than in the Standard Model which motivates the searches for some specific final states. 
The following is a list of the CMS searches reviewed in this paper.

\begin{itemize}
  \item Search for a neutral Higgs boson decaying to a pair of b-quarks and produced in association with 
    at least one additional b-quark ($pp\rightarrow bH+X,H\rightarrow b\bar{b}$)
  \item Search for a neutral Higgs boson decaying to a pair of $\tau$ leptons 
    ($pp\rightarrow H+X,H\rightarrow \tau\tau$)
  \item Search for a neutral Higgs boson decaying to a pair of opposite sign muons ($pp\rightarrow H+X,H\rightarrow \mu^{+}\mu^{-}$)
\end{itemize}

Each of the three analysis mentioned above have their own sensitivity, background composition and event selection, the 
fully description and results are reported in independent analysis notes and publications, in this paper only general 
details about each analysis are discussed. In the search for a neutral Higgs boson in a final state with b-quarks~\cite{hbb} 
the most important background contribution comes from QCD multi-jet events, the analysis was divided into two categories: 
the "all-hadronic" and the "semi-leptonic", in the latest one the presence of a non-isolated muon from a b-jet is 
required per event, both signatures use data-driven techniques to estimate the background contribution and the analysis strategy is 
to search for an excess of events in the invariant mass distribution of the two leading b-jets. 
In the search with a final state with two $\tau$ leptons~\cite{htautau} three final states were considered: e$\tau_{h}$,$\mu \tau_{h}$ and $e\mu$, each lepton candidate was required to be isolated since $\tau$'s from 
Higgs bosons are commonly isolated from the rest of the events activity, the events were divided in two categories: 
the b-tag and the non-btag and the analysis strategy was to search for an excess of events in the di-tau invariant mass distribution.
For the search of a neutral Higgs boson decaying to a pair opposite sign muons the main background is the Drell-Yan process, 
there are less important sources of background like $t\bar{t}$ and $W^{\pm}W^{\pm}$, more details about the background estimation 
methods for every search is presented in table~\ref{tab:mssmbkg}

\begin{table}[!htb]
\begin{center}
\begin{tabularx}{\textwidth}{|p{1.3in}|c|p{1.1in}|X|}  \hline
Search   &    Channel     &   Main Background   &  Estimation Method \\ \hline
\multirow{2}{*}{$pp\rightarrow bH,H\rightarrow b\bar{b}$}  & all-hadronic    & QCD multi-jet & Templates~from~a~double-btag-sample in data\\ \cline{2-4}
                                                             & semi-leptonic & QCD multi-jet &b-tagging probabilities and nearest-neighbor-in-parameter-space technique\\ \hline
$pp\rightarrow H\rightarrow \tau\tau$  & $e\tau_{h}$,$\mu \tau_{h} e\mu$   & $Z\rightarrow\tau\tau$ &  Using a sample of $Z\rightarrow \mu\mu$ events where reconstructed muons are replaced by simulated taus ("embedding") \\  \hline
$pp\rightarrow H\rightarrow \mu^{+}\mu^{-}$  & $\mu^{+}\mu^{-}$   & Drell-Yan   & Background and signal shapes extracted from a fit to the di-$\mu$ invariant mass in data  \\ \hline 

\end{tabularx}
\caption{Summary of the main background processes for MSSM searches and description of the estimation method.}
\label{tab:mssmbkg}
\end{center}
\end{table}

\begin{figure}[!hbp]
  \begin{center}
    \includegraphics[width=0.49\textwidth]{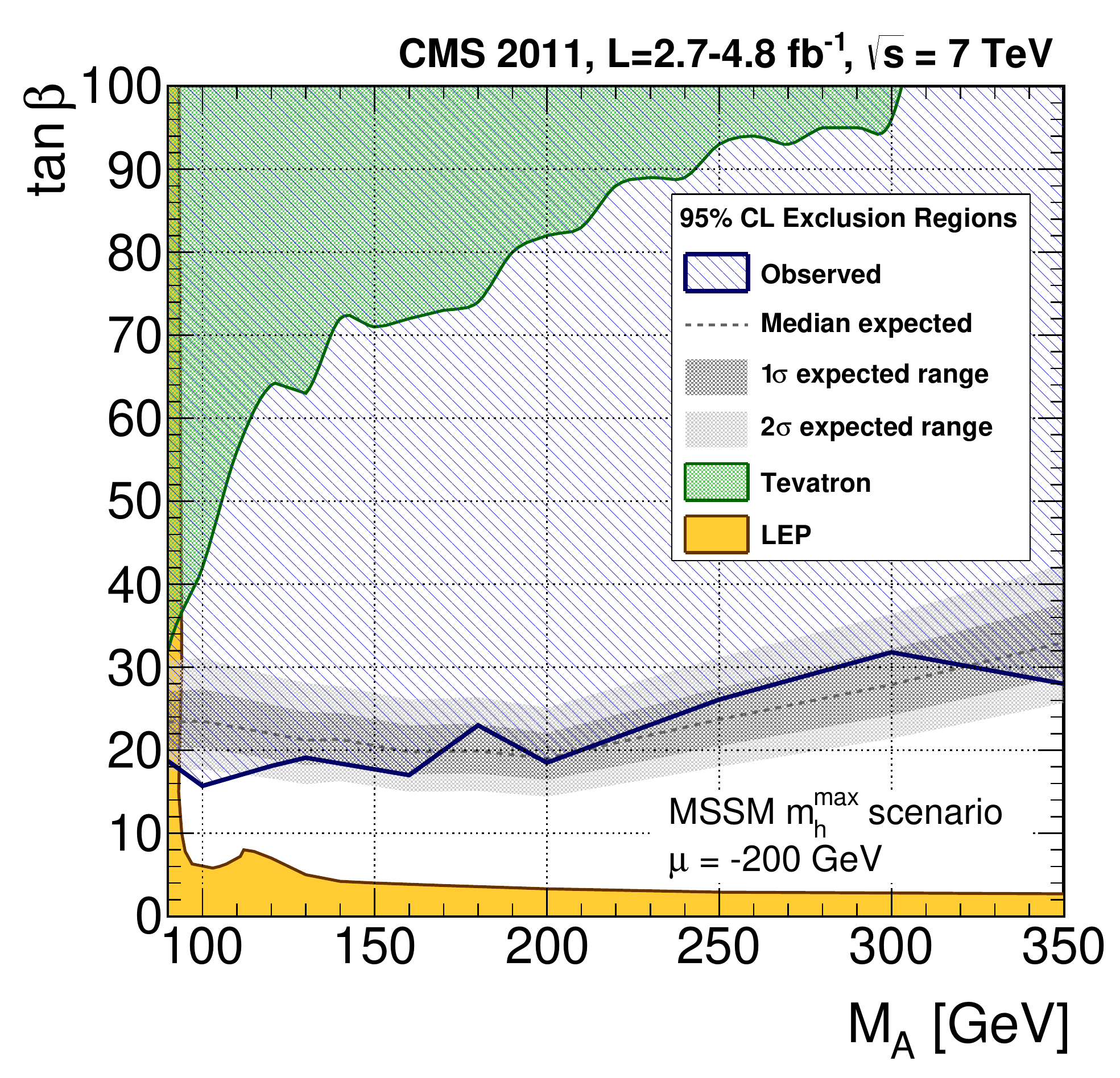}
    \includegraphics[width=0.49\textwidth]{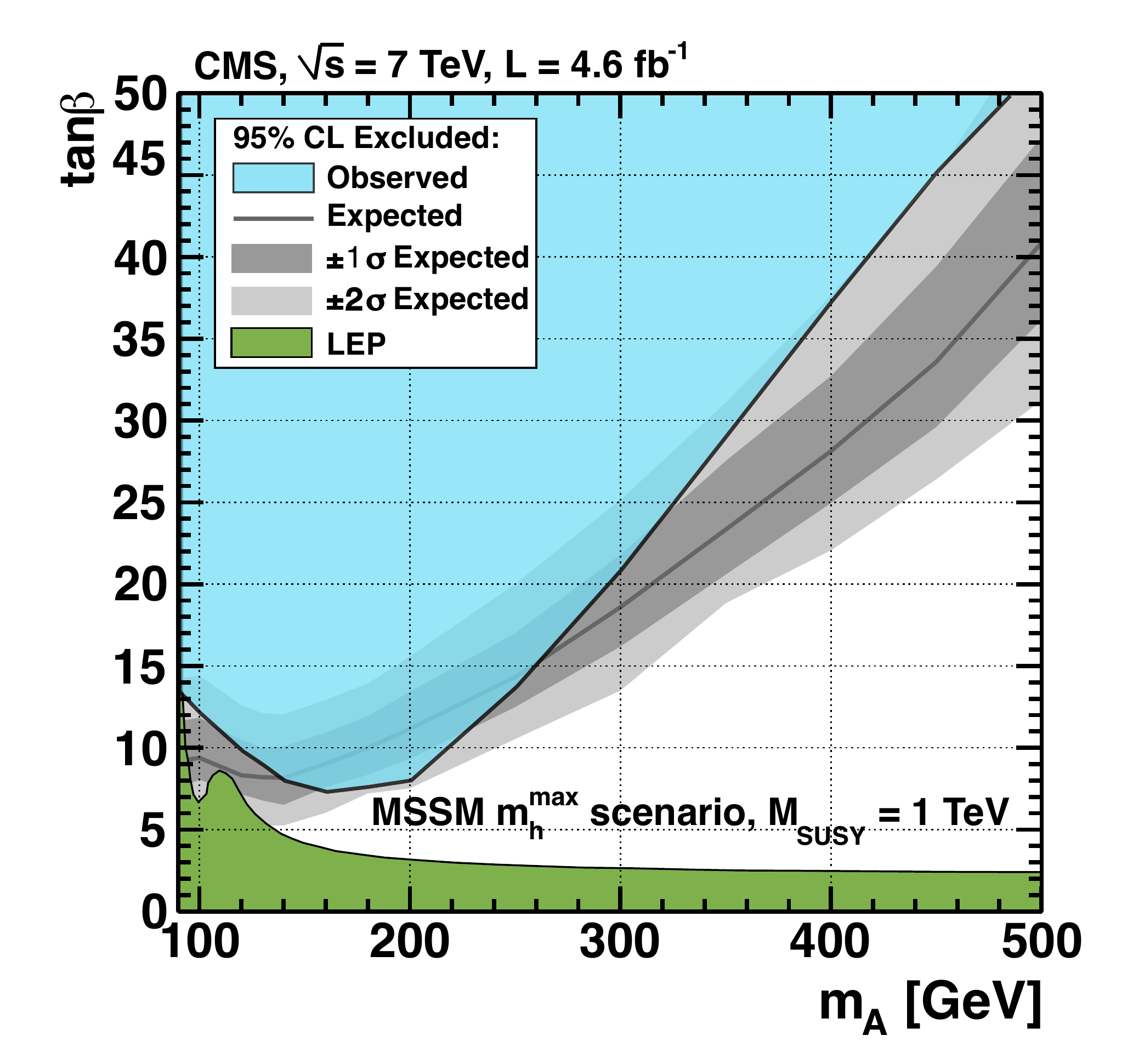}
    \caption{Left: Observed upper limits at 95\% confidence level on $tan{\beta}$ as a function of $m_{A}$ for the search 
      of a neutral Higgs boson decaying to a pair of b-quarks for the combined "all-hadronic" and "semi-leptonic" results, 
      previous exclusion regions from LEP and 
    Tevatron in the multi-b jet channel are overlaid.  Right: Region in the parameter space of $\tan{\beta}$ 
versus $m_{A}$ excluded at 95\% CL in the context of the MSSM for the search of a neutral Higgs boson decaying to a pair of tau leptons. The expected one and two standard 
    deviations ranges and the observed 95\% CL upper limits are shown together with the observed exclusion region.}
    \label{fig:limits_mssm}
    \end{center}
\end{figure}

Neither of the three searches reported an excess of events over the expected Standard Model background and limits are set 
on $tan(\beta)$ as a function of $m_{A}$ as shown in Figure~\ref{fig:limits_mssm}.

\section{NMSSM searches in CMS}

The Next-to-Minimal Supersymmetric Model (NMSSM)~\cite{nmssm} was introduced to remove some of the fine-tuning used in MSSM, 
several reasons justify the study of this extended theory and some of them are the following:
\begin{itemize}
  \item Naturally solves in an elegant way the so-called $\mu$ problem, where $\mu$ is a dimensional parameter, in MSSM this parameter has to be adjusted by hand to a value at the electroweak scale, it is desirable to have an extension in which this $\mu$-term can be generated dynamically.
  \item In MSSM the Higgs sector is highly restricted (lightest Higgs is always the Standard Model like). An extended Higgs sector may relax this restriction and lower the experimental bounds (in NMSSM the lightest Higgs boson can be as light as 1~GeV)
\end{itemize}
In the MSSM an additional gauge singlet is introduced which generates the $\mu$-term dynamically, that is, an 
effective $\mu$-term arises spontaneously and the adjustment by hand drops out. This is surely the main 
motivation for the NMSSM and may justify the price to pay, which is the introduction of an additional gauge-singlet superfield. The particle content in the bosonic part of the singlet results in two additional Higgs bosons giving a total of seven compared with the five Higgs bosons 
in MSSM.


Searches for a light Higgs boson in the context of NMSSM have been explored in CMS, two of them are reviewed in this paper:
\begin{itemize}
  \item Search for a light pseudo-scalar Higgs boson ($a$) decaying to a pair of muons of opposite sign ($a\rightarrow \mu^{+}\mu^{-}$)
  \item Search for a non-Standard Model Higgs boson decaying to a pair of new light bosons which further decay to a pair of opposite sign muons $h\rightarrow 2a\rightarrow 4\mu$
\end{itemize}

\begin{figure}[!hbp]
  \begin{center}
    \includegraphics[width=0.60\textwidth]{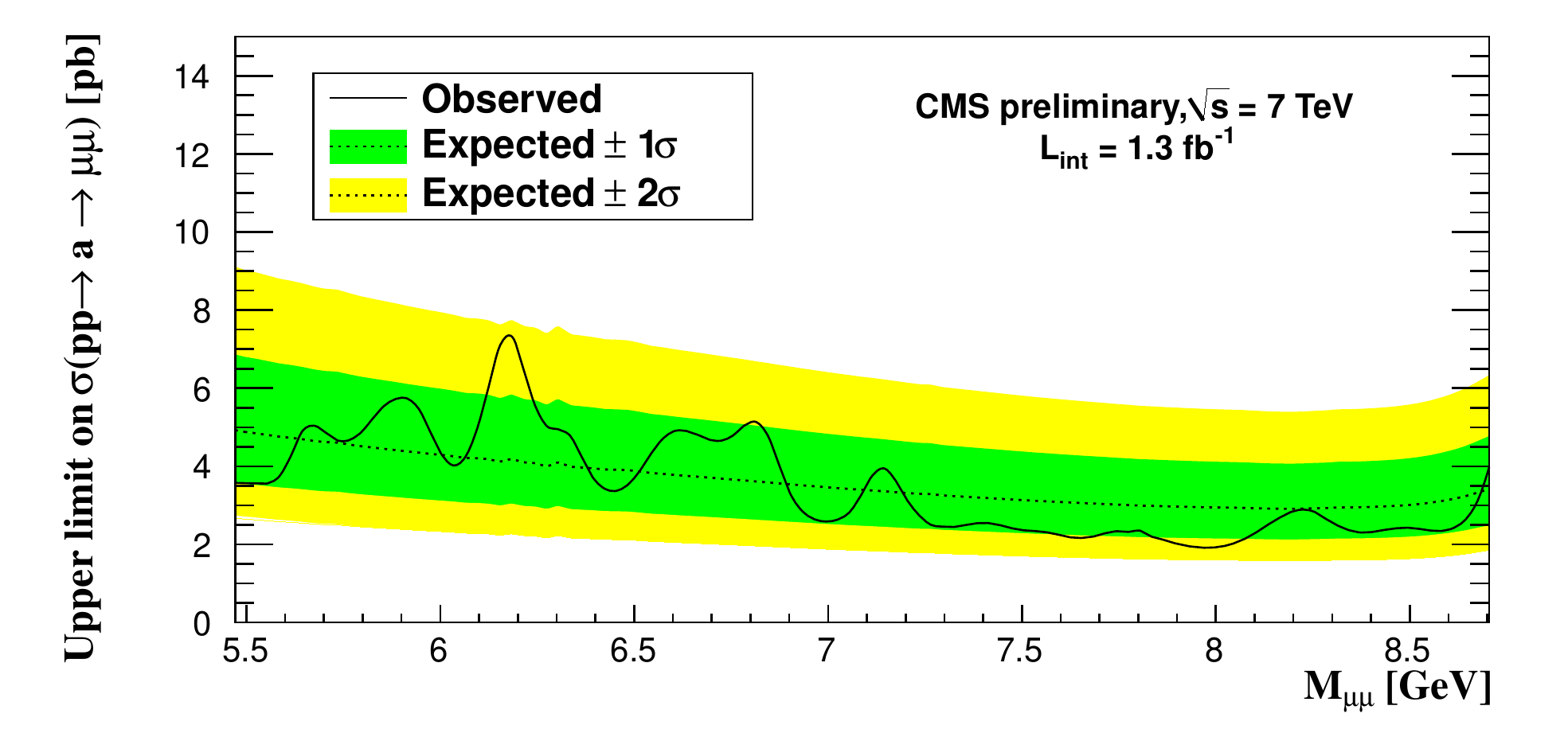}
    \includegraphics[width=0.60\textwidth]{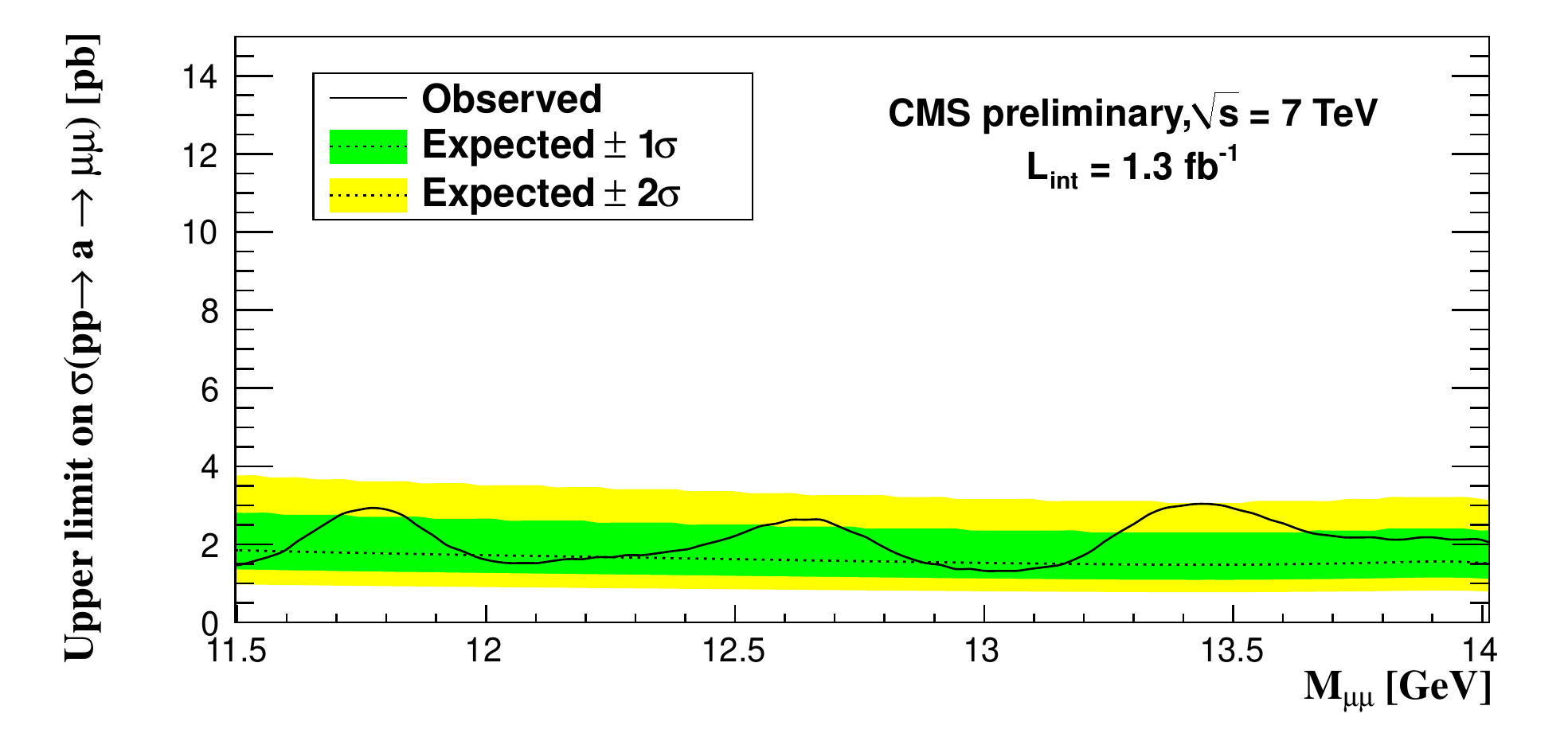}
    \caption{Upper limits at 95\% C.L. on $\sigma\times B(pp\rightarrow a \rightarrow \mu^{+}\mu^{-})$ in mass 
    range 1 (top) and mass range 2 (bottom) including systematic uncertainties. The dotted lines correspond to 
    the expected limits, and the bands correspond to 1-and 2-$\sigma$ level uncertainties on the expected limits.}
    \label{fig:limits_amumu}
    \end{center}
\end{figure}

\begin{figure}[!hbp]
  \begin{center}
    \includegraphics[width=0.48\textwidth]{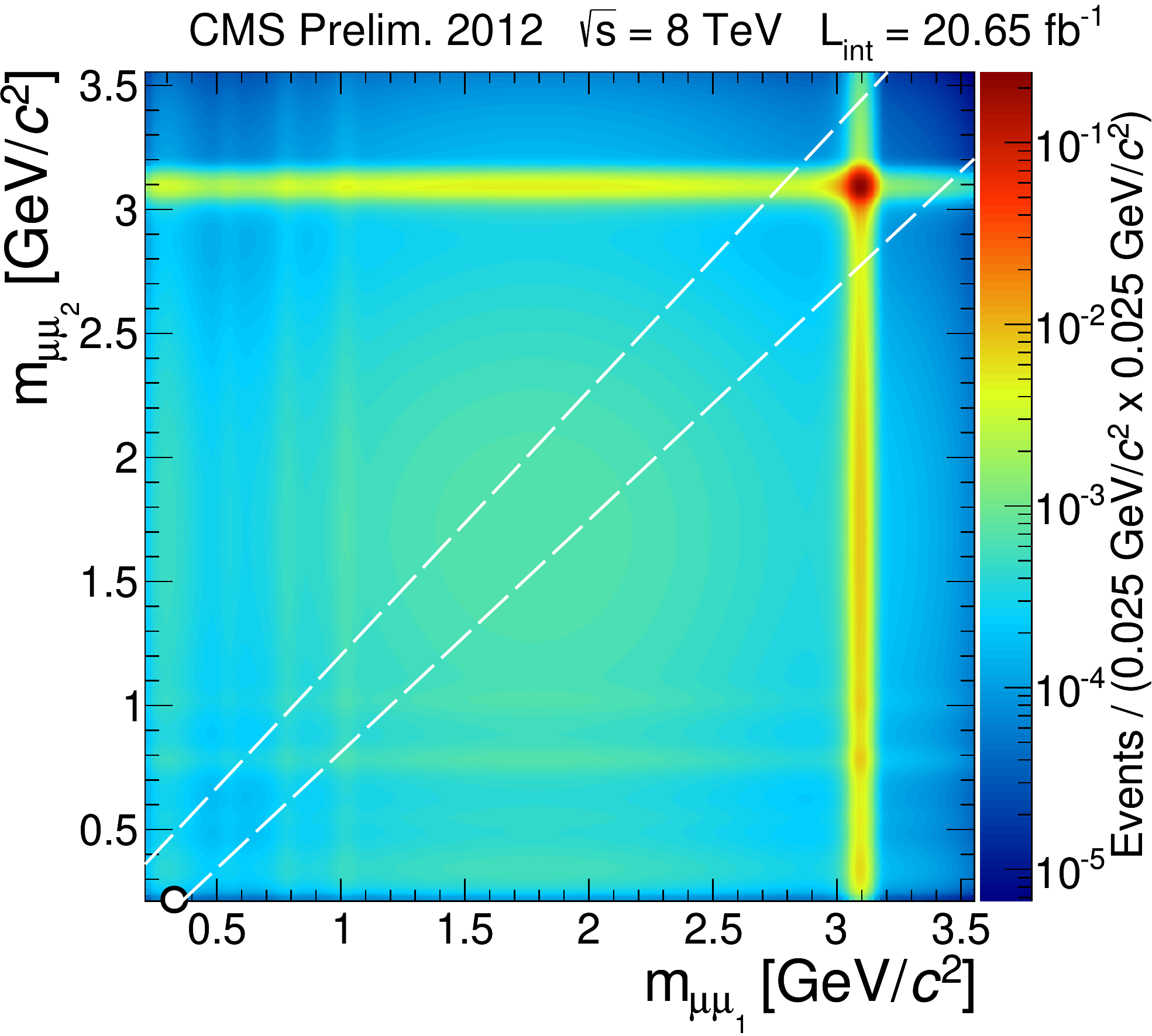}
    \includegraphics[width=0.48\textwidth]{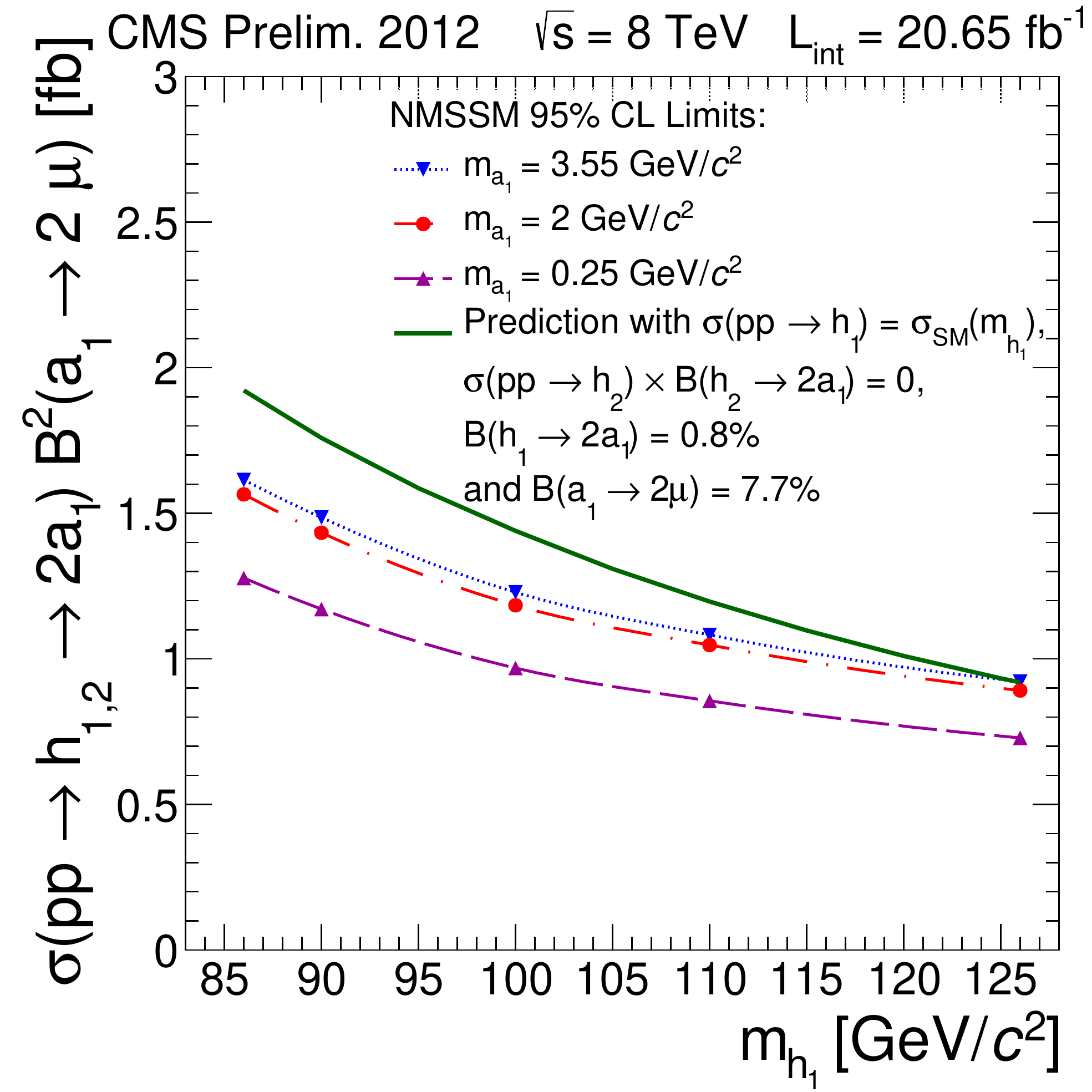}
    \caption{Left: $m_{\mu\mu_{1}}$ vs $m_{\mu\mu_{2}}$ for the isolated di-muon system showing the one event 
      in the data (shown as empty circle) which survived all selection requirements, the intensity of the shading indicates the 
background expectation which is a sum of the $b\bar{b}$ and the direct $J/\psi$ pair production contribution. 
Right: The $95\%$ CL upper limits as functions of $m_{\text{h}_1}$, for the NMSSM case, on $\sigma(pp \to \text{h}_{1,2} \to 2 \text{a}_1) \times \mathcal{B}^2(\text{a}_1 \to 2 \mu)$ with $m_{\text{a}_1}=0.25GeV$ (dashed curve), $m_{\text{a}_1}=2GeV$ (dash-dotted curve) and $m_{\text{a}_1}=3.55GeV$ (dotted curve). As an illustration, the limits are compared to the predicted rate (solid curve) obtained using a simplified scenario with $\sigma (pp \to \text{h}_1)=\sigma_\mathrm{SM}(m_{\text{h}_1})$, $\sigma (pp \to \text{h}_2) \times \mathcal{B}(\text{h}_{2} \rightarrow 2 \text{a}_1) = 0$, $\mathcal{B}(\text{h}_1 \to 2 \text{a}_1) = 0.8\%$, and $\mathcal{B}(\text{a}_1 \to 2\mu) = 7.7\%$. }
    \label{fig:4mu}
    \end{center}
\end{figure}

The search for a light Higgs boson decaying to two opposite sign muons~\cite{amumu} is constrained from previous measurement from 
Babar in the region $m_{a}<m_{\Psi (1s)}$ while for the region $m_{a}>m_{\Psi (3S)}$ the Tevatron and the 
LHC have sensitivity. This channel is characterized by a large cross section but a small branching ratio to muons, the invariant 
mass range was divided in 
two separate regions, the first one from 5.5 to 8.8~GeV and the second one from 11.5 to 14~GeV, the region in 
between was excluded due to the abundant contribution of bottonium resonances, no significance excess of events were found in either of these 
regions and limits were set on the production cross section multiplied by the Branching ratio to muons as shown in Figure~\ref{fig:limits_amumu}
The second analysis is a search for a pair of new light bosons in the context of NMSSM, each light bosons decay 
to a pair of muons giving a final topology of a pair of di-muons with compatible mass ($m_{di-\mu_{1}} \sim m_{di-\mu_{2}}$). The main backgrounds are the production of a pair of b-quarks ($b\bar{b}$) and the direct production of double $J/\psi$, the $b\bar{b}$ background is estimated directly from data by selecting a $b\bar{b}$ enriched sample with 
events with one di-muon plus additional "orphan" muon, in this control region a background template is taken from 
the fit to the di-muon invariant mass, such template is validated in events with two di-muons. The double-$J/\psi$
background is estimated using Monte Carlo expectations and applying corrections from measurements in data. 
After all the event selection only one event survived and it is compatible with Standard Model background expectation, 
limits were set as a function of one of the non-Standard Model Higgs boson mass, considering different benchmark signal scenarios 
as shown in Figure~\ref{fig:4mu}.

\section{Summary and Outlook}

The CMS collaboration has a broad program for Exotic Higgs searches, some of them have already produced results 
analyzing the available data from CMS 2011 and 2012 data campaigns. 
In this paper some of the searches in the context of MSSM and NMSSM theories were reviewed, 
in neither of them was found a significant excess of events over Standard Model expectations, therefore 
limits were set as a function of the parameters of each extended theory, such limits improve the coverage from 
previous experiments like Tevatron and LEP. The CMS collaboration is looking forward to accumulate more data in the second 
phase of the experiment (beyond 2015) to confirm or reject the consistency of the new particle with the predicted Standard Model 
Higgs boson.

\vspace{-0.2in}
\Acknowledgments
The corresponding author is supported by the Qatar National Research Fund (QNRF) under project NPRP-5-464-1-080

\end{document}

%% file: econfmacros.tex
\def\beq{\begin{equation}}
\def\eeq#1{\label{#1}\end{equation}}
\def\eeqn{\end{equation}}

\def\beqa{\begin{eqnarray}}
\def\eeqa#1{\label{#1}\end{eqnarray}}
\def\eeqan{\end{eqnarray}}

\let\bar=\overbar

\def\Dslash{\not{\hbox{\kern-4pt $D$}}}
\def\dslash{\not{\hbox{\kern-2pt $\del$}}}

\def\msb{{\bar{\ssstyle M \kern -1pt S}}}

